\documentclass[aps,prl,twocolumn,groupedaddress,showpacs,byrevtex]{revtex4}
\usepackage{graphicx}    % needed for including graphics e.g. EPS, PS

\begin{document}
\title{Antiferromagnetic Spatial Ordering in a Quenched One-dimensional Spinor Gas }
\author{A. Vinit, E. M. Bookjans, C. A. R. S\'a de Melo and C. Raman}
\email {chandra.raman@physics.gatech.edu}
\affiliation{School of Physics, Georgia Institute of Technology, Atlanta, Georgia 30332, USA}
\date{\today}

\begin{abstract}
We have experimentally observed the emergence of spontaneous antiferromagnetic spatial order in a sodium spinor Bose-Einstein condensate that was quenched through a magnetic phase transition.  For negative values of the quadratic Zeeman shift, a gas initially prepared in the $F = 1, m_F=0$ state collapsed into a dynamically evolving superposition of all 3 spin projections, $m_F = 0,\pm 1$.  The quench gave rise to rich, nonequilibrium behavior where both nematic and magnetic spin waves were generated.  We characterized the spatiotemporal evolution through two particle correlations between atoms in each pair of spin states.  These revealed dramatic differences between the dynamics of the spin correlations and those of the spin populations.

\end{abstract}

\pacs{03.75.Mn,67.85.De,67.85.Fg,67.85.Hj}
\maketitle

Dynamics far from equilibrium occur ubiquitously in nature.  A prime example can be found close to a symmetry breaking phase transition.  A rapid passage through the critical point can quickly nucleate spatial inhomogeneities that subsequently evolve slowly with time.  Such behavior typifies physical systems of vastly different microscopic origin, for example, rapid cooling of the early universe following inflation \cite{kibble1976,zure96,vilenkin1994}, the chiral symmetry breaking phase transition in heavy-ion collisions \cite{rajagopal1993,berdnikov2000} and sudden cooling of solid-state magnets below the Curie temperature \cite{bray94}.  The nature of the pattern formation is of fundamental interest to uncover its universal features \cite{polkovnikov2011,lamacraft2012}.  For example, the temporal evolution of correlation functions can yield critical exponents of the phase transition \cite{calabrese2006}.  Understanding how entanglement and correlations flow in a many-body system is also relevant for enabling quantum computation \cite{bose2007}.

\begin{figure}[htbp]
\includegraphics[width= 1.1 \columnwidth]{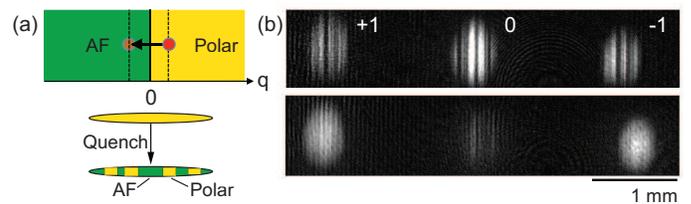}
%\vspace{-0.27 in}
\caption{(Color online).  Quantum quench dynamics.  (a) Instantaneous change in quadratic Zeeman energy $q$ brings the gas through a phase transition between polar ($m_F = 0$) and antiferromagnetic ($m_F = \pm 1$) ground states at $t=0$.  Elongated BEC (yellow ellipse) subsequently develops one-dimensional spin domains.  (b) Time-of-flight Stern-Gerlach images show the spontaneous axial spin structures at a hold time of 48 ms (upper image) while by 2.5 s (lower image) the system has relaxed to an antiferromagnetic configuration free of domains.  Residual fringes in the latter image are experimental artifacts due to the probe light.}\label{fig:ONE}
%\vspace{-0.2 in}
\end{figure}

The spin degrees of freedom of a Bose gas offer exciting possibilities for the exploration of non-equilibrium phenomena \cite{stamper-kurn2012,matuszewski2008,matuszewski2010}.  In this work we have realized a quantum quench in the laboratory by crossing through a magnetic phase transition in an antiferromagnetic, $F = 1$ sodium spinor Bose-Einstein condensate (BEC), as shown in Figure \ref{fig:ONE}.  We have captured the stochastic dynamics of this quench through spatially resolved measurements of two particle spin correlations.  The spin-dependent mean-field Hamiltonian,
\[  H_{sp} =  c_2 n \langle \hat{\bf F}\rangle ^2 + q \langle \hat{F}_z^2 \rangle  \]
contains a competition between intrinsic magnetic interactions and the coupling to an external magnetic field.  The latter is parameterized by the quadratic Zeeman shift $q$, which we rapidly switched from positive to negative values to initiate the quench.  Similar correlation measurements have been performed recently for quenched superfluids without spin \cite{cheneau2012,hung2012}.  In the above formula, $n$ is the mean particle density and $c_2$ is the strength of spin-dependent interactions, which can be either positive (for antiferromagnets) or negative (for ferromagnets).  $\hat{\bf F}$ and $ \hat{F}_z$ are the vector spin 1 operator and its $\hat{z}$ projection, respectively.  A variety of quantum phases can be realized depending on the value of $q$ \cite{stamper-kurn2012}.  In $F = 1$ $^{87}$Rb, where $c_2 < 0$, altering $q$ suddenly through a phase boundary resulted in spontaneous transverse magnetization patterns \cite{sadler06}.  

The case $c_2 > 0$ has been much less studied.  Here it is nematic, rather than magnetic, order that is the principal degree of freedom \cite{ho98,zhou01,mukerjee06,barnett2010}, which profoundly affects the nature of the spatial patterns and their dynamics \cite{muel04}.  Early work on antiferromagnetic gases principally explored  instabilities in two spin components, where nematic order played no role \cite{mies99meta}.  In our experiment, we observe spatial dynamics after tuning $q$ through a phase transition at $q = 0$ using microwave magnetic fields, a transition first observed in our earlier work \cite{bookjans2011}.  Related work on $F = 2$ $^{87}$Rb BEC, for which $c_2 > 0$, triggered pattern formation at a specific wavelength by preparing an unstable initial state \cite{kronjager2010}.  Our work explores complementary dynamics, where passage through a phase transition generates fluctuations in the nematic order on all length scales rather than just a specific wavevector as in \cite{kronjager2010}.  Thus the path to equilibrium involves the collective dynamics of {\em many} modes, which is our focus.

We used time-of-flight Stern-Gerlach (TOF-SG) observation to record the one-dimensional spatiotemporal pattern formation in all 3 spin components, $m_F = 0,\pm 1$, with high resolution.  This method works only for extreme aspect ratio traps \cite{cast96}.  Importantly, we also resolved the spin density {\em fluctuations}, which revealed the crucial role played by the spin-mixing interaction between atom pairs, $2 (m_F = 0)  \leftrightarrow (m_F = 1) + (m_F = -1)$.  This interaction leads to entanglement between spins \cite{pu2000,lucke2011,bookjans2011b}, and to density correlations since the fluctuations are simply related by $\Delta n_0 = -2 \Delta n_{-1} = -2 \Delta n_{+1}$.  In our experiment we show these to initially develop {\em locally}, and then to evolve {\em globally}.  We have spatially resolved the full spin-density correlation matrix of a quantum quench, to our knowledge for the first time.  In thermal equilibrium these density correlations directly determine the compressibility matrix \cite{sademelo2011,sademelo1991}.  

Optically confined Bose-Einstein condensates in the $m_F = 0$ state were prepared in a static magnetic field of $B_x = 100 $ mG in a manner described in our earlier work \cite{bookjans2011}.  Transverse magnetic fields and field gradients were compensated to within $ 5$ mG and $ 0.6$ mG/cm, respectively.  The peak density $n_0 = 5 \times 10^{14} \rm{cm}^{-3}$ and axial Thomas-Fermi radius $R_x = $ 340 $\mu$m were measured to an accuracy of 5\%, from which we determined the spin-dependent interaction energy $c_2 n_0 = h \times 120$ Hz.  The axial and radial trapping frequencies were 7 and $470$ Hz, respectively, accurate to 10\%.  The radial Thomas-Fermi radius, $R_\perp = $ 5 $\mu$m, was small enough such that only axial spin domains could form \footnote{The energy available from the quench, $h \times 4 $ Hz, is smaller than the transverse excitation frequency which we estimate to be $h \times $50 Hz from a 2-dimensional box model.}.  The measured temperature was 400 nK, close to the chemical potential of 360 nK.  

We rapidly switched $q$ from $q_i = h \times +2.8$ Hz $> 0$ to a final value $q_f = h \times -4.2$ Hz $< 0$ at $t=0$.  A quantum phase transition divides the polar ($m_F = 0$) ground state from an ``antiferromagnetic'' one ($m_F = \pm 1$ superposition) at $q=0$, as shown in Figure \ref{fig:ONE}.  The dynamical evolution from one to the other state is  governed by the spin-dependent interaction term $\propto c_2$.  $q$ was switched by turning on a far-off resonance microwave ``dressing'' field tuned to near the $F = 1 \rightarrow F = 2$ ground state hyperfine resonance.  The value of $q_f$ was controlled through the AC Stark shift by adjusting the microwave power \cite{gerbier2006}.    

\begin{figure}[htbp]
\includegraphics[width= \columnwidth]{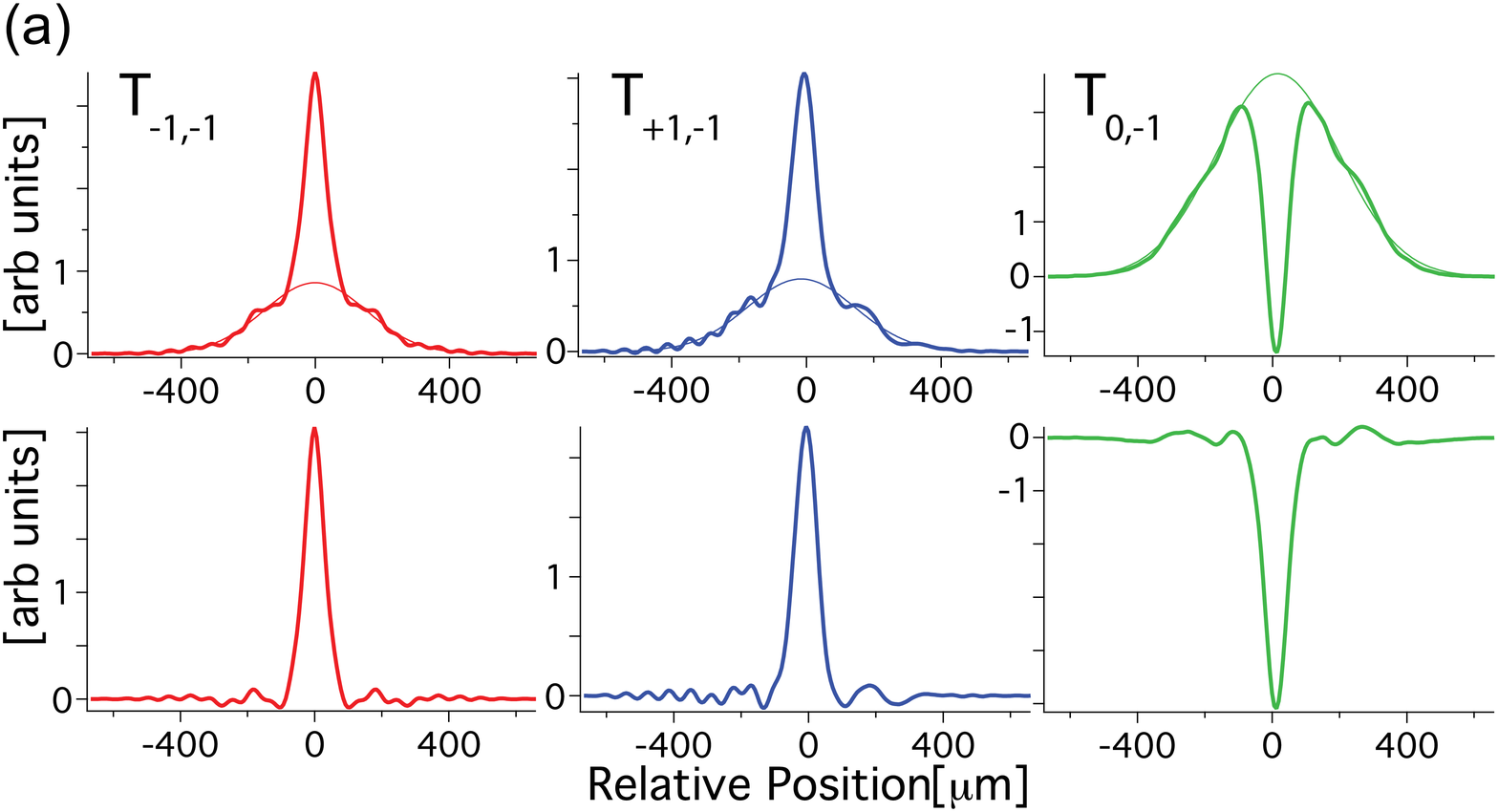}
\vspace{-0.1 in}
\includegraphics[width=\columnwidth]{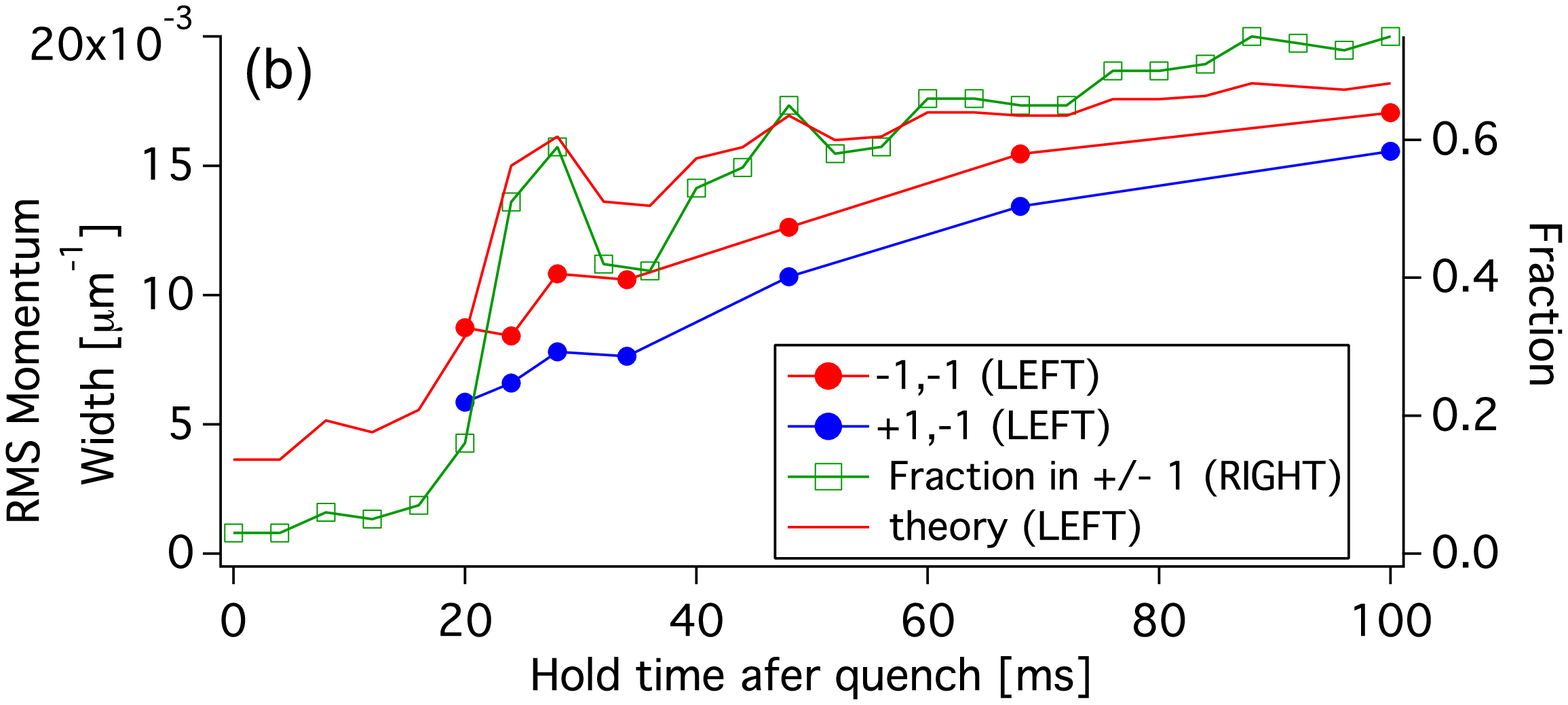}
%\vspace{-0.27 in}
\caption{(Color online.) Nonlocal correlations and spin domain dynamics. (a) Top row shows $T_{i,j}(\delta) = \int S_{i,j}(x,x+\delta) dx $ for various spin pairs and relative position $\delta$.  Bottom row shows the same data with the background caused by atom number variations subtracted.  (b) Dynamics of the fraction of atoms in $\pm 1$ (green squares, right axis), the RMS momentum width determined from the Fourier transform of $T_{-1,-1}$ and $T_{-1,+1}$ (red and blue circles, respectively, left axis) and from a theoretical prediction (solid red line, left axis).}
\label{fig:FOUR}
%\vspace{-0.2in}
\end{figure}

Since $q_i,|q_f|\ll c_2 n_0$, the gas remained very close to the phase transition point at all times.  Nonetheless the change triggered a rapid instability in the $m_F=0$ condensate.  Through spin exchange collisional interactions we observed that the $m_F = \pm 1$ population fraction, $f_{pm}$, increased rapidly near $t= 20$ ms (see Figure \ref{fig:FOUR}b).  This was accompanied by oscillations and a slower period of relaxation over $> 100 $ ms toward an apparent equilibrium.  We observed a 30\% atom loss over 100 ms due to off-resonant microwave excitation.

After holding the gas for a time $t$, the initial spin density fluctuations were amplified, and ultimately, macroscopic, one-dimensional spatial domains were observed to have formed (see Figure \ref{fig:ONE}b).  We then diagnosed the axial spin distribution using a TOF-SG sequence \cite{sten98spin} consisting of 2 ms TOF, 4 ms pulsed SG field and 24 ms additional TOF to separate the 3 spin states, followed by a 50 $\mu$s absorption imaging pulse on the $F = 2 \rightarrow F' = 3$ transition of the D2 line.  Due to the extreme 70:1 aspect ratio the TOF axial distribution remained very close to its {\em in-situ} value.  Each image was Fourier filtered to remove spurious interference fringes with wavelength between 60-62 $\mu$m created by the probe light, and summed over the radial direction to obtain one-dimensional atom density distributions.  We took the center of these distributions ($x = 0$) to be the maximum of the average density profile for 30 runs at each hold time $t$, after smoothing by a 200 $\mu$m moving average filter.  

\begin{figure}[htbp]
\includegraphics[width= 0.9\columnwidth]{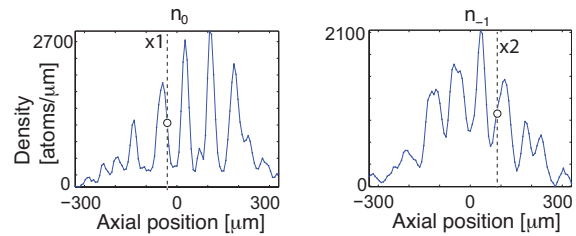}
\caption{(Color online.) Sample profiles $n_0(x)$ and $n_{-1}(x)$ measured on a single shot at $t = 68$ ms.}\label{fig:TWO} 

%\vspace{-0.2in}
\end{figure}

Figure \ref{fig:TWO} shows measured spin density profiles, $n_0(x)$ and $n_{-1}(x)$, for a single shot of the experiment at a hold time of 68 ms.  Domains as small as 13 $\mu$m half-width at half-maximum can be seen.  This is 4\% of the Thomas-Fermi radius, and is comparable to our imaging resolution, $d \approx $ 10 $\mu$m \footnote{$d$ was determined by both the 6.6 $\mu$m camera pixel size and the blurring along the imaging axis due to the finite 300 $\mu$m radius of the expanded cloud.  The latter imposed an effective resolution of 7.5 $\mu$m due to depth of focus considerations.}.  Domains smaller than $d$ would have expanded to a size $\ge d$, and thus cannot be distinguished by the time-of-flight method from larger scale features.  The energetics of the quench, however, suggest that domains no smaller than $\hbar/(2 \sqrt{M q_f}) = 5 \mu$m are likely to appear.  

While single images contained multiple spin domains, the average over many runs showed no structure, indicating that the formation and dynamics were stochastic in nature.  Therefore, we quantified the data using two-particle correlations, specifically computing the two-point covariance function (TPCF) for the density fluctuations $\Delta n$ of a spin pair $i,j$, 
\[S_{i,j} (x_1,x_2,t) \equiv \langle \Delta \hat{n}_i (x_1,t) \Delta \hat{n}_j(x_2,t) \rangle\]
where $\langle...\rangle$ refer to ensemble averages of fluctuations about the sample mean, taken over $30$ runs of the experiment.  Averages over a smaller sample size yielded similar results.  We further analyzed these data in terms of both the local correlation function ($x_1 = x_2$) as well as non-local effects ($x_1 \neq x_2$).  We discuss both these results in turn below.

\begin{figure}[htp]
\includegraphics[width = 0.9\columnwidth]{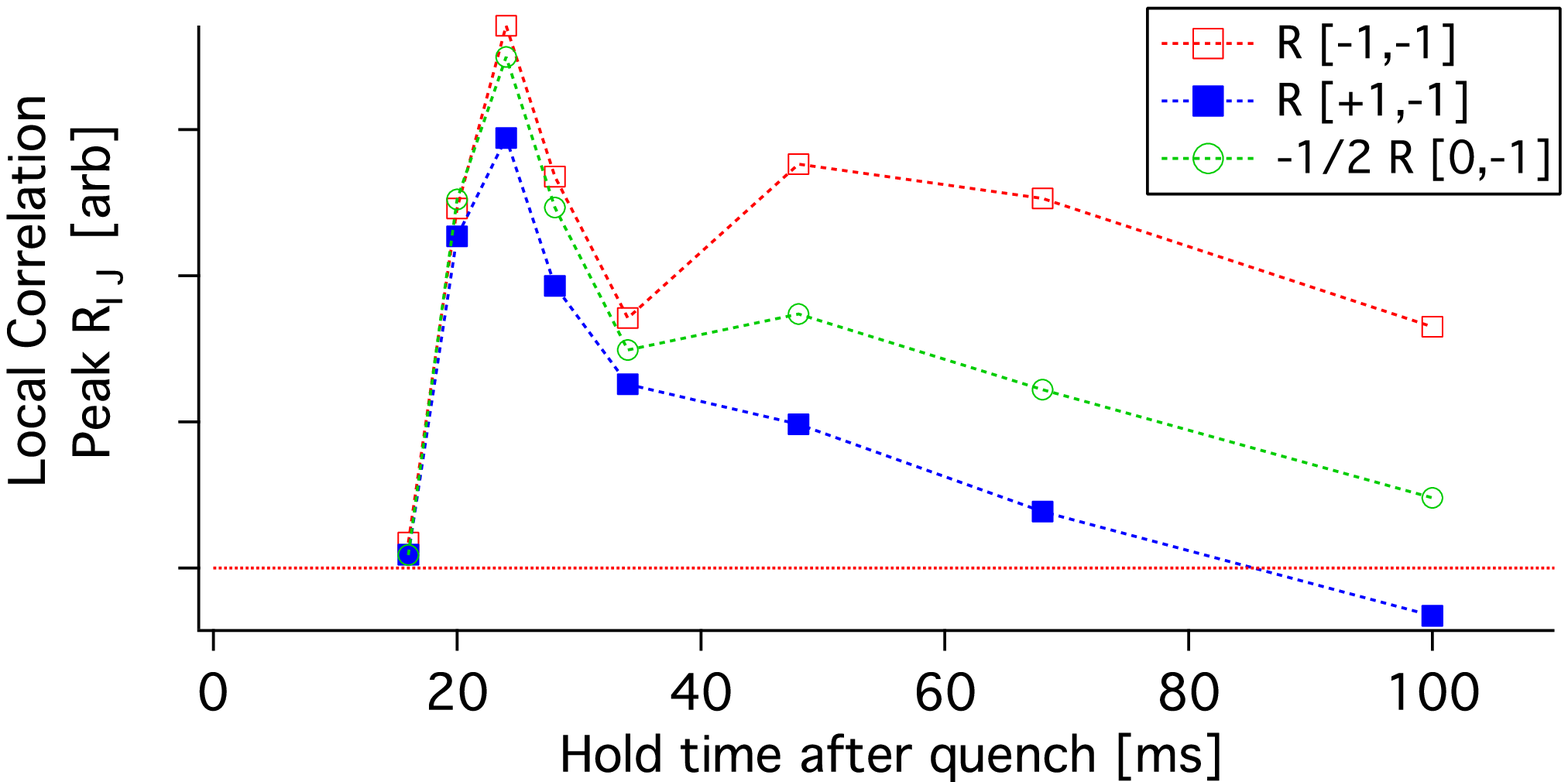}
\includegraphics[width = 0.9\columnwidth]{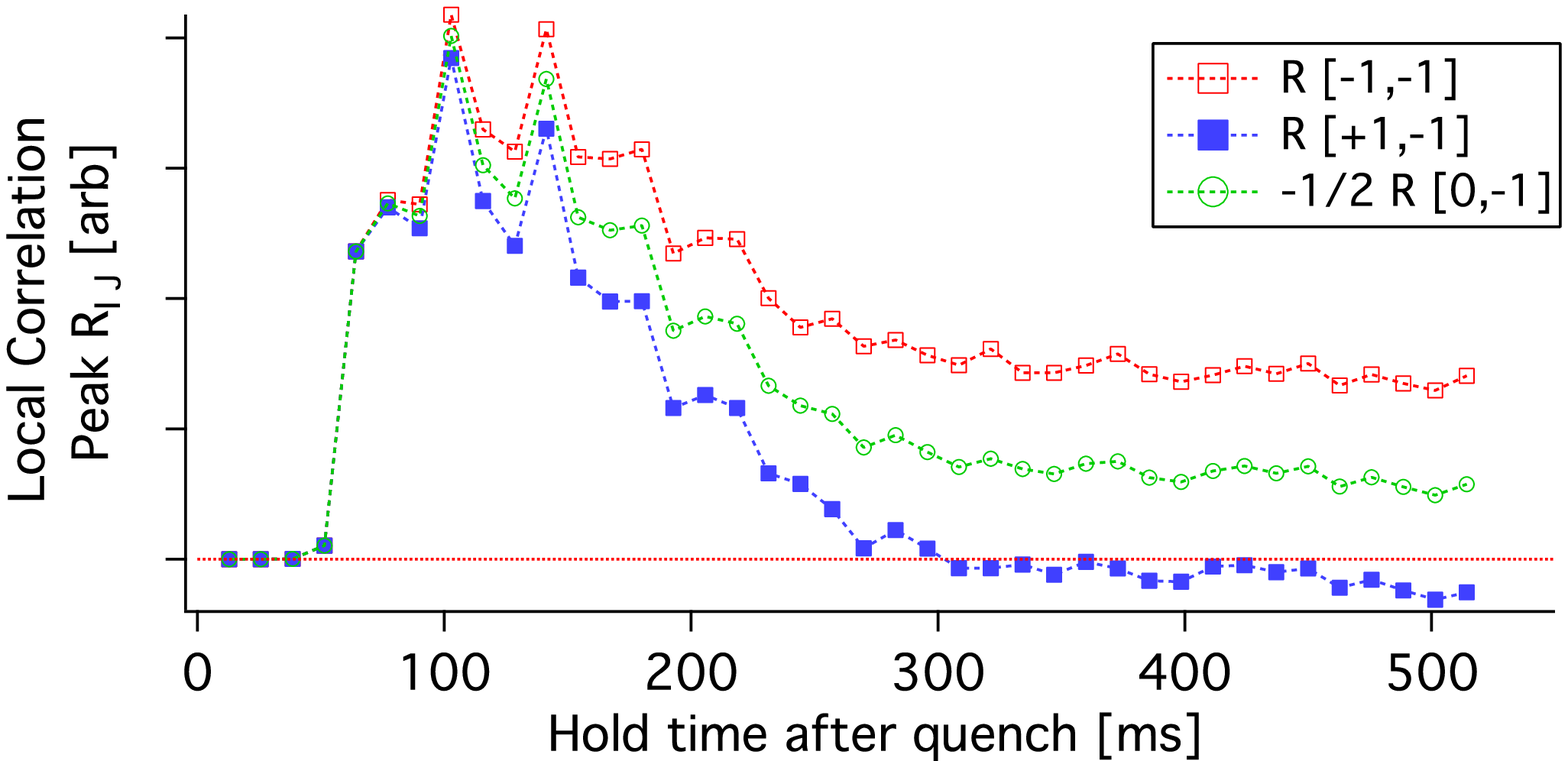}

%\vspace{-0.27 in}
\caption{(Color online.) Correlation of atom pairs after the quench at $t = 0$.  (a) Measured local density correlations  $R_{ij}(t) = \int dx \langle \Delta \hat{n}_i(x,t) \Delta \hat{n}_j(x,t) \rangle$ for spin states $(i,j) =  (-1,-1)$ and $(+1,-1)$ versus time.  Also shown is the $(0,-1)$ correlation scaled by a factor  $-\frac{1}{2}$ as described in the text.  (b) Correlation functions determined from numerical simulations. The horizontal dashed line marks zero correlation.  }\label{fig:THREE}
%\vspace{-0.25in}
\end{figure}

Figure \ref{fig:THREE}a shows the measured time evolution of the local pair correlation function, $R_{ij}(t) = \int dx \langle \Delta \hat{n}_i(x,t) \Delta \hat{n}_j(x,t) \rangle$.  This graph is a central result of this work.  Noise due to total atom number fluctuations was subtracted from the data as described later in the manuscript.  For short times, $t \lesssim  35$ ms, all 3 spin states exhibit correlations whose absolute value rises from zero to a maximum value.  This is a result of the local spin mixing process discussed earlier, which predicts that the ratio $R_{1,-1}/R_{-1,-1} = 1.0$ and $R_{0,-1}/R_{-1,-1} = -2.0$.  Our measured peak values of the ratios are $0.9$ and $-2.0$, respectively.  Thus our data indicate that although the system is outside the single-mode regime \cite{stamper-kurn2012}, spin mixing generates nearly the maximum local correlations that are possible.

What is equally interesting is that for $t \gtrsim 35$ ms the data all diverge from one another.  The diagonal correlations, $R_{-1,-1}$, appear to reach a steady state of 50 \% of their peak value.  This is the result of saturation of the parametric gain due to the finite number of $m_F = 0$ atoms, which causes fluctuations in the amplified modes to be suppressed \footnote{They are not, however, reduced to zero, since the same spin state must always be correlated with itself at the same location.}.  By contrast, the off-diagonal data, $R_{-1,+1}$, decay sharply.  The $R_{0,-1}$ data has been scaled by the factor $-\frac{1}{2}$ to allow for comparison of all 3 curves, and exhibits an intermediate behavior between $R_{-1,-1}$ and $R_{+1,-1}$.  

The initial, positive correlation between $\pm 1$ atoms is indicative of nematic spin wave excitations, which are small oscillations in the order parameter $N_{zz} = n_{+1}(x,t) + n_{-1}(x,t) - 2 n_0(x,t)$.  These excitations occur naturally for $F = 1$ spinor condensates with $c_2 > 0$, since $m_F = 0$ regions phase separate from $m_F = \pm 1$ \cite{sten98spin}.  As time evolves, distinct regions in the condensate can communicate with one another through the propagation of these spin waves.  The decay of $R_{+1,-1}$ can therefore be viewed as a destructive intererence between waves originating from different points in space.  Interference effects have also been recently reported in quenched scalar superfluids, leading to oscillations in the density-density correlation function \cite{hung2012}.  

Accompanying this decay of $R_{+1,-1}$ is a growth of {\em magnetic} spin wave excitations, whose order parameter is $M_z(x,t) = n_{+1}(x,t)- n_{-1}(x,t)$.  The local magnetization variance is ${\rm var}[M_z(x,t)] = S_{+1,+1}+ S_{-1,-1}-2S_{-1,+1}$, which becomes non-zero for imperfect correlation between the two spins, $S_{-1,+1}(x,t) < S_{+1,+1}(x,t),S_{-1,-1}(x,t)$, as we observed experimentally.  Our data in Figure \ref{fig:THREE}a show that $R_{+1,-1}$ reaches zero at $\approx 85$ ms, after which the two spins even become negatively correlated.

In order to understand our data theoretically, we have performed numerical simulations of the 3 coupled spinor Gross-Pitaevskii (GP) equations.  Assuming a BEC initially at zero temperature, we obtained the initial wavefunction for the $m_F = 0$ component numerically.  Vacuum noise in the $m_F = \pm 1$ states was simulated as classical noise according to the Truncated Wigner approximation (TWA) \cite{blakie2008,polkovnikov2010,sau2009,barnett2011}.  Thirty separate simulations allowed us to compute the expectation value of quantum mechanical observables as statistical averages over different random initial conditions.  Vacuum modes with wavelength less than $\xi_{spin}$ are not expected to contribute to the spin instability.  Therefore, we imposed a cutoff energy of $c_2 n_0$, which resulted in $\simeq $700 virtual particles, while the condensate contained $5 \times 10^6$ particles, similar to the experimental conditions.  The  results were relatively insensitive to this cutoff as expected for an exponential amplification process.  For example, magnifying the noise by a factor of 10 changed the time at which the $\pm 1$ fraction $ = 0.5$ by only 20 \%.  

Figure \ref{fig:THREE}b shows the numerically obtained correlation functions, which shows good qualitative agreement with the experimentally observed trends.  Similar to the experimental data, $R_{+1,-1}$ has dropped below zero in the simulation after $\approx 300$ ms.  The overall dynamics were faster in the experiment by approximately this factor, $300/85 = 3.5$, which we cannot presently account for.  Possible effects include residual $\pm 1$ population in the initial state caused by thermal or technical fluctuations that would speed up the instability.  In addition, atom loss (in our case due to off-resonant microwave excitation) can cause decoherence \cite{bookjans2011b}.

The simulations show that even in the absence of thermal or technical noise, vacuum fluctuations can induce a decay of correlations, a somewhat counterintuitive result.  For any particular realization of the experiment, the vacuum noise of the two bosonic fields, $\hat{\psi}_{+1}$ and $\hat{\psi}_{-1}$, is uncorrelated.  Tiny differences in the initial conditions become amplified by the spin mixing instability, resulting in final wavefunctions $\psi_{\pm 1}(t = t_F)$ that differ macroscopically, i.e., by much more than just the vacuum noise.  By contrast, identical initial conditions for $\psi_{\pm 1}$ (an unphysical situation) resulted in identical time evolution and no decay, consistent with the symmetry of the GP equations \cite{stamper-kurn2012}.

We conclude by quantifying the non-local dynamics.  Figure \ref{fig:FOUR} shows a one-dimensional reduction of the covariance map $S_{ij}$ in terms of relative coordinate $\delta$, $T_{i,j} (\delta,t) = \int \langle \Delta n_i(x,t) \Delta n_j(x + \delta,t) \rangle d x$, for spin pairs $(-1,-1)$, $(+1,-1)$ and $(0,-1)$, at $t = $20 ms after the quench.  A narrow peak, either positive or negative, can be seen near $\delta = 0$.  A smooth, positive background lies underneath it, primarily caused by atom number variations from one shot to the next.  The latter lead to correlated density fluctuations throughout the cloud \footnote{The area under the background was observed to largely follow the overall populations during the quench.}.  The wings of the data appeared to fit well to a Gaussian distribution.  When subtracted, the data revealed the {\em intrinsic} fluctuations, which are plotted underneath, showing a sharp, local correlation peak surrounded by oscillations.  The height of the peak was determined by an additional Gaussian fit, and constitutes the data of Figure \ref{fig:THREE}.  

The oscillations signify a dynamically evolving non-local order induced by the rapid quench, which we characterized by computing the Fourier transforms, $T_{-1,-1}(k)$ and $T_{+1,-1}(k)$.  These showed an increase in high spatial frequency components with time after the quench.  Figure \ref{fig:FOUR}b shows this trend as a growth in the second moment of the Fourier transforms, $k_{rms} \equiv \left (\int_0^\infty k^2 T dk/\int_0^\infty T dk \right )^{1/2}$.  Higher order effects in the SG expansion are responsible for the slight magnification of $k_{rms}$ for the $(-1,-1)$ relative to $(+1,-1)$.  

While a rapid quench is expected to lead to coarsening of the domains on long timescales \cite{bray94,hohenberg1977}, our data show the opposite trend within the time window 0-100 ms.  This suggests that it might be an intermediate phase of evolution that precedes coarsening.  In fact, energy conservation requires the kinetic energy to grow with the $m_F = \pm 1$ populations, since they reduce the quadratic Zeeman energy.  Using $\hbar^2 k_{rms}^2/(2M)$ from the correlation functions as an estimate of the mean kinetic energy, we obtain $ \frac{k_{rms}(t)}{2 \pi} = \sqrt{-2 M q_f f_{\pm 1}}/h^2 = f_{\pm 1}(t)^{1/2}/(47 \mu {\rm m})$.  This formula is plotted as a solid line Figure \ref{fig:FOUR}b using our measurements, and shows the same trend as our data.  It consistently lie above our measurements, most likely due to the fact that we do not account for variations in the phase of the order parameter nor the interaction energy contained in the finite width domain boundaries.

In summary, we have measured the time evolution of spin correlations in a quenched spinor BEC.  Our data reveal rich non-equilibrium dynamics where both nematic and magnetic spin density waves are spontaneously generated.  Future work will explore possible universal signatures in the decay of correlations \cite{calabrese2006}.

We thank A. Polkovnikov, R. Hippolito, M. Vengalattore, and P. Goldbart for useful discussions.  This work was supported by DoE grant No.\ DE-FG02-03ER15450 and NSF grant No.\ 1100179.

%\bibliographystyle{apsrev}
%\bibliography{References}

\end{document}